\documentclass[reprint,prl,twocolumn,superscriptaddress,nofootinbib,floatfix]{revtex4-1}
\usepackage{graphicx,bm,color,amsmath,amssymb,caption,float}

\newcommand{\ba}{\begin{eqnarray}}
\newcommand{\ea}{\end{eqnarray}}
\newcommand{\be}{\begin{equation}}
\newcommand{\ee}{\end{equation}}
\newcommand{\bi}{\begin{itemize}}
\newcommand{\ei}{\end{itemize}}

\newcommand{\la}{\lambda}

\newcommand{\sa}{\sigma}

\newcommand{\n}{\nabla}

\newcommand{\ra}{\rightarrow}

\newcommand{\LF}{\left(}
\newcommand{\RF}{\right)}

\newcommand{\4}{\frac{1}{4}}

\newcommand{\Fc}{\mathcal{F}}

\begin{document}

\title{Behavior of the Newtonian potential for ghost-free gravity  and singularity free gravity}

\author{James Edholm}
\affiliation{Consortium for Fundamental Physics, Lancaster University, Lancaster, LA1 4YB, United Kingdom}

\author{Alexey S. Koshelev}
\affiliation{Departamento  de F\'isica and Centro  de  Matem\'atica  e 
Aplica\c c\~oes,  Universidade  da  Beira  Interior,  6200  Covilh\~a, 
Portugal}
\affiliation{Theoretische Natuurkunde, Vrije Universiteit Brussel, and
 International Solvay Institutes, Pleinlaan 2, B-1050 Brussels, Belgium }

\author{Anupam Mazumdar$^{1}$}
\noaffiliation

\begin{abstract}
In this paper we show that there is a universal prediction for the Newtonian potential for a specific class of {\it infinite derivative, ghost-free, quadratic curvature} 
gravity. We show that in order to make such a theory {\it ghost-free} at a perturbative level, the Newtonian potential always falls-off as $1/r$ in the infrared
limit, while at short distances the potential becomes non-singular. We provide examples which can potentially test the scale of gravitational non-locality up to $0.004$~eV.

\end{abstract}
\maketitle

\section{Introduction}

Einstein's theory of General Relativity (GR) has  passed successfully through innumerable tests from small scales to large scales~\cite{Will}.
One of its predictions, of the existence of gravitational waves, has recently been confirmed by the advanced Laser
Interferometer Gravitational-Wave Observatory (LIGO), which has observed a transient gravitational-wave (GW) signal and tested the reliability 
of GR~\cite{Abbott:2016blz}.  In all these examples, in the infrared (IR), the theory matches the 
Newtonian fall of $1/r$ potential. In spite of these great successes, the theory of GR is incomplete in the 
ultraviolet (UV), the classical solutions of GR exhibit black hole and cosmological type singularities, and at 
a quantum level the theory is not UV finite. GR definitely requires modifications in the UV; the question is what 
kind of corrections in the UV one would expect, which would make the theory well behaved in the classical and in quantum sense,
and possibly resolve the short distance singularities.

For a massless graviton, in $4$ dimensions, all the interactions in the UV can in {\it principle} be captured by
incorporating higher derivatives allowed by the {\it diffeomorphism-invariance}. For instance,
it is well-known that higher derivatives can ameliorate the UV behaviour, i.e.
$4th$ derivative gravity is renormalisable, but at a cost of introducing a {\it ghost} term in the spin-2 component of a 
graviton propagator~\cite{Stelle}. Indeed, the presence of ghosts can lead to a destabilising of the classical vacuum, therefore rendering the theory unpredictable at both classical and at a quantum level.

Recently, the issue of ghosts has been addressed  in the context of {\it quadratic} gravity - in order to make the theory generally covariant and
{\it ghost-free} at the perturbative level, one would require infinite derivatives~\cite{Biswas:2011ar,Biswas:2005qr}. Indeed, these {\it infinite} derivatives would modify the graviton propagator. However, if we capture the roots of these infinite derivatives by the {\it exponential of an entire function}, then there will be no new degrees of freedom propagating in spacetime other than the massless transverse and traceless graviton, since such modification of the graviton propagator would not 
introduce any new pole.

It has been demonstrated that these infinite derivatives with a graviton propagator modified by the exponential of an entire function can indeed 
soften the quantum UV behaviour~\cite{Tomboulis,Siegel,Deser:1986xr,Jack:1987pc,Modesto,Talaganis:2014ida}. 
Furthermore, such a prescription also removes the cosmological Big Bang singularity~\cite{Biswas:2005qr,Biswas:2010zk}, 
and blackhole type singularity in a static limit~\cite{Biswas:2011ar}, and in the dynamical context~\cite{Frolov:2015bia}, in a linearised limit. One intuitive way to understand this is due to the fact that infinite derivatives render the gravitational
interactions non-local~\cite{Tomboulis,Talaganis:2014ida}. This non-locality also introduces an inherent new scale in $4$ dimensions, i.e. 
$M\leq M_p\sim 2.4\times 10^{18}$~GeV. Furthermore, an intriguing connection can be established between the
gravitational entropy~\cite{Wald}, and the propagating degrees of freedom in the spacetime. The gravitational entropy for ghost-free, infinite gravity
does not get a contribution from the UV, but only from the Einstein-Hilbert action~\cite{Conroy:2015wfa}, and follows strictly the area - law for entropy
for a Schwarzschild's black hole.

The aim of this paper is two-fold: first we show that for a wide class of infinite derivative theories of gravity which are {\it ghost-free}, it is possible to recover {\it not only} the 
$1/r$ fall of the Newtonian potential in a static limit in the IR, but also to ameliorate the short distance behaviour in the UV limit. Second, we wish 
to put a bound on the scale of non-locality, i.e. $M$, from the current table-top experiments from deviation of Newtonian gravity.

\section{Quadratic Curvature Gravitational Action }

Let us first start by discussing the properties of GR in $4$ dimensions. The linearised GR can be quantised around the Minkowski background, 
which is described by $2$ massless degrees of freedom. The transverse and traceless components of the 
graviton propagator in $4$ dimensions can be recast in terms of the spin projector operators, which involves the tensor 
${\cal P}^{(2)} $, and only one of the scalar components, i.e. ${\cal P}_s^{(0)}$~\cite{VanNieuwenhuizen1973}:
\be\label{pi}
\Pi(k^2) \sim \frac{1}{k^2} \LF \mathcal{P}^{(2)}-\frac{1}{2}\mathcal{P}_{s}^{(0)} \RF\,,
\ee
where $k^{\mu}$ is the $4$-momentum vector, where we have suppressed the spacetime indices.

In fact, in~\cite{Biswas:2016etb,Biswas:2011ar} it has been shown that around the Minkowski background, in $4$ dimensions, the most general {\it quadratic order}
torsion-free and parity invariant gravitational action which can be made {\it ghost-free} can be written in terms of the Ricci-scalar, $R$, the symmetric traceless tensor, 
$S_{\mu\nu}=R_{\mu\nu}-\frac{1}{4}Rg_{\mu\nu}$, and $C_{\mu\nu\alpha\beta}$ is the Weyl tensor. It is sufficient to study the quadratic order action - which captures
${\cal O}(h^2)$ terms around the Minkowski background, i.e. $g_{\mu\nu}=\bar g_{\mu\nu}+h_{\mu\nu}$, where $\bar g_{\mu\nu}$ is the Minkowski background, and $h_{\mu\nu}$ are the
excitations, in order to find the graviton propagator. The $S$-tensor vanishes on maximally symmetric backgrounds (Minkowski or (anti)-de Sitter)~\cite{Biswas:2016etb}~\footnote{The original action was written in terms of $R_{\mu\nu}$ and $R_{\mu\nu\lambda\sigma}$ in~\cite{Biswas:2011ar}. However there is no loss of generality in expressing the action as Eq.~(\ref{ligoproperaction}), see~\cite{Biswas:2016etb}. See also~\cite{Deser:1986xr,Jack:1987pc,Woodard}, where ghost conditions have been studied in the context of string theory.}, therefore the full action can be written as: 
\begin{eqnarray}
S = \int d^4x \sqrt{-g}\left[\frac{M_P^2}{2} R
+\frac{\lambda}2\Bigg(R\Fc_1(\Box)R \right.\nonumber \\
\left. +S_{\mu\nu}\Fc_2(\Box)S^{\mu\nu}+ 
C_{\mu\nu\la\sa}\Fc_{3}(\Box)C^{\mu\nu\la\sa}\Bigg)\right]\,,
\label{ligoproperaction}
\end{eqnarray}
where $M_P^2$ is the Planck mass, and $\lambda$ is a dimensional coupling accounting for the higher curvature modification, 
and the $\Fc_i$ are Taylor expandable (i.e. analytic) functions of the covariant d'Alembertian~\cite{Biswas:2011ar}, i.e.
%
   $     \Fc_i(\Box)=\sum_{n=0}c_{i_n}\Box^n/M^{2n} $,
%
where $M$ is the scale of non-locality. 

The equations of motion of this action have been worked out in~\cite{Biswas:2013cha}.
As we shall show now, this class of infinite derivative theory 
indeed provides a unique platform to study departure from GR in future {\it table-top} experiments~\cite{Adelberger}.

\section{Universality of the Newtonian potential}
 Physical excitations of this action, Eq.~(\ref{ligoproperaction}), around Minkowski background have been studied very well. This can 
 be computed by the second variation of the action, using
       $ g_{\mu\nu}=\bar g_{\mu\nu}+h_{\mu\nu} $.
A quick computation can be made by employing the covariant mode decomposition of the metric \cite{D'Hoker}:
\begin{equation}
        h_{\mu\nu}={\tilde h}_{\mu\nu}+\bar\n_{\mu}A_{\nu}+\bar\n_\nu A_\mu+(\bar \n_{\mu}\bar\n_{\nu}-\4
\bar g_{\mu\nu}\bar \Box)B+\4 \bar g_{\mu\nu}h,\label{ligodecomphabh}
\end{equation}
where $\tilde h_{\mu\nu}$ is the transverse and traceless spin-2 excitation, $A_\mu$ is a transverse vector field, and $B,~h$ are two scalar 
degrees of freedom which mix. Upon linearization around maximally symmetric backgrounds, 
the vector mode and the double derivative scalar mode vanish identically, and we end up only with $\tilde h_{\mu\nu}$
 and $\phi=h-\Box B$~\cite{Biswas:2016etb}. 
Performing necessary computations (which are indeed straightforward around Minkowski as all derivatives commute), one gets~\cite{Biswas:2016etb}:
\begin{widetext}
\begin{equation}
\begin{split}
&\delta^2 S(\tilde h_{\nu\mu})=\int
dx^4\sqrt{-\bar g}\frac12\tilde h_{\mu\nu}\bar\Box a(\bar\Box)\tilde h^{\mu\nu},~~~~~~a(\bar\Box)=1+\frac\lambda{M_P^2}\bar\Box\left({\Fc}_2(\bar\Box)+2\Fc_3\left(\bar \Box\right)\right)\\
&\delta^2 S(\phi)=-\int
dx^4\sqrt{-\bar g}\frac12 \phi\bar\Box c(\bar\Box)\phi,~~~~~~~~c(\bar\Box)=1-\frac\lambda{M_P^2}\bar\Box\left(6\Fc_1(\bar \Box)+\frac1{2}{\Fc}_2\left(\bar \Box\right)\right)
\end{split}\label{ligotensor}
\end{equation}
\end{widetext}
for the tensor component (where the field was rescaled by $M_P/2$ to become canonically normalised), and the scalar component 
(where the field was rescaled by $M_P\sqrt{3/32}$ to be canonically normalised), respectively.

The full graviton propagator can then be written using a similar method to~\cite{VanNieuwenhuizen1973}, 
barring the suppressed indices~\footnote{In~\cite{VanNieuwenhuizen1973}, 
the authors imposed $6$ projection operators to decompose spin $2$ and spin $0$ component of the propagator, 
here we have employed a slightly different technique to decompose the $10$ metric degrees of freedom.}~\cite{Biswas:2013cha,Biswas:2013kla,Biswas:2016etb}:
\begin{eqnarray}\label{modpi}
        \Pi(k^2)&=&\frac{{\cal P}^{(2)}}{k^2a(-k^2)}+\frac{{\cal P}^{(0)}}{k^2(a(-k^{2})-3c(-k^{2}))}\,,
\end{eqnarray}
where ${\cal P}^{(2),~(0)}$ are the spin projection operators \cite{VanNieuwenhuizen1973}. Note that the graviton propagator 
has two unknown functions $a(k^2)$ and $c(k^2)$, where all the information about the infinite derivatives is hiding, see~\cite{Biswas:2011ar,Biswas:2013kla,Biswas:2013cha}
for an alternative way of deriving the graviton propagator, Eq.~(\ref{modpi}), and related discussion on form factors. It is possible
that $a(\bar{\Box})$ and $c(\bar{\Box})$ are not uniquely defined under field redefinitions~\cite{Deser:1986xr,Jack:1987pc,Siegel}, but this issue is beyond the scope of this paper.

In order to reduce the graviton propagator to that of GR, one method is to assume that $ a(\bar\Box)= c(\bar\Box)$. In the IR limit then both $a(k^2\to 0)= 1,~~c(k^2\to 0)=1$, 
such that Eq.~(\ref{modpi}) reduces to Eq.~(\ref{pi}).
In this limit the theory would match exactly GR's predictions in the IR, but would lead to modification in the UV. The entire modification can be summarised 
by one unknown function $a(\bar\Box)$, which constrains the functions such that, see for instance~\cite{Biswas:2013cha}: $$12 \Fc_1(\bar\Box)+6\Fc_2(\bar\Box)+4\Fc_3(\bar\Box)=0.$$
In order that the propagator have no poles except the massless graviton at $k=0$, we require that $a(\bar\Box)$ and $(a(\bar\Box)-3c(\bar\Box))$ must not contain any
zeros.
 This way the propagator, Eq.~(\ref{modpi}), will not contain any 
 extra degrees of freedom propagating in the space-time other than the massless graviton with 2 helicity states. One possible choice is to 
 assume $a(\bar\Box)$ is the exponential of an {\it entire function}. This choice makes sure that in spite of infinite derivatives, there exist {\it no ghosts} at the perturbative level
 for a quadratic curvature gravity Eq.~(\ref{ligoproperaction}). One such example will be~\cite{Siegel,Biswas:2005qr,Biswas:2011ar}
\begin{equation}
        a(\bar\Box)=c(\bar\Box)=e^{-\bar\Box/M^2}.
        \label{tsprl}
\end{equation}
 This choice  guarantees that in the UV the theory is softened, as for $k\ra \infty$, $a(-k^2)=c(-k^2) =e^{k^2/M^2}$
 suppresses the propagator in the UV, i.e. $\Pi (k^2) \ra 0$ in Eq.~(\ref{modpi}), while $k\ra 0$ yields the pure 4D GR propagator.
  
 
 Our aim in this paper will be to generalise this to any entire function $\tau(-k^2)$,
such that in the momentum space we have:
\begin{equation}
        a(-k^2)=c(-k^2)=e^{-\tau(-k^2/M^2)}\,.
\end{equation}
The computation of the Newtonian potential, i.e. $\Phi(r)$, for the simplest choice, when $\tau(-k^2/M^2)=-k^2/M^2$ as in Eq.~(\ref{tsprl}), was done already in 
~\cite{Biswas:2005qr}, and the result is
\begin{equation}
        \Phi(r)\sim -\frac{\mu}{M_P^2r}\sqrt{\frac{\pi}2}\mathrm{erf}(Mr/2),
        \label{newtonerf}
\end{equation}
where $\mu$ is the mass of a $\delta$-source. This potential is finite near $r\approx0$ and decays as $1/r$ at distances above the non-locality 
scale, i.e. $r \gg M^{-1}$. The tests of $1/r$ fall of Newtonian gravity has been tested in the laboratory up to $5.6 \times 10^{-5}$m~\cite{Kapner:2006si}, which implies that for the 
scale of non-locality should be bigger than $M >0.004~{\rm eV}$. Indeed, we know very little about the gravitational interaction above this limit!
The cornerstone of this computation is the {\it sine Fourier} transform
\begin{equation}
        f(r)=\int_{-\infty}^{+\infty}\frac{dk}ke^{\tau(-k^2)}\sin(kr),
        \label{newtonsine}
\end{equation}
where 
\ba
        \Phi(r)=  -\frac{\mu}{4\pi^2 M^2_p} \frac{f(r)}{r}.
\ea
When we consider the simplest choice, $\tau=-k^2/M^2$, the function $f(r)$ indeed gives an $\mathrm{erf}$-function.

We now set out to prove that the leading behaviour of the potential at small distances, $r$, away from the source
is always given by: $\Phi\approx\Phi_0+{\cal O}(r)$, where $\Phi_0$ is constant irrespectively of the form of an {\it entire function} $\tau(k^2)$, as long as 
it does not introduce any extra pole other than the massless graviton.

\section{Generalisations of the Entire Function}

\begin{figure}[!htbp]\includegraphics[width=91mm]{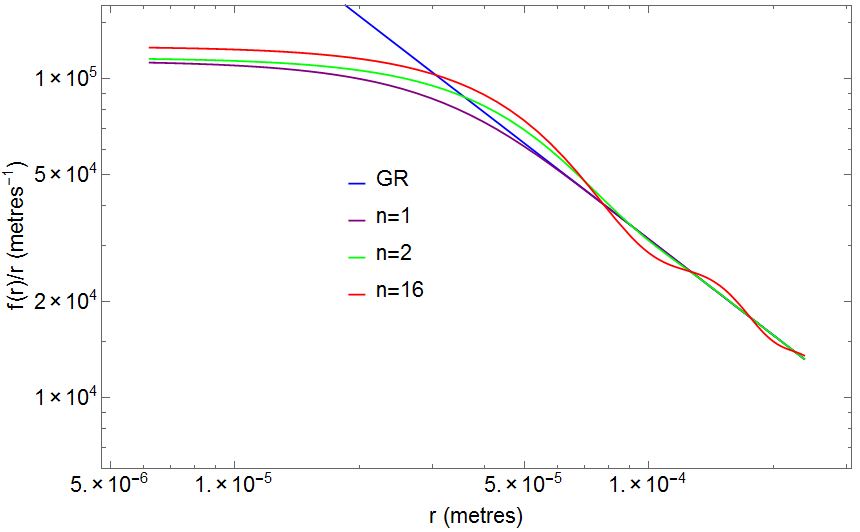}
\caption{We plot  $f(r)/r$ vs $r$ for different $n$  for Eq.~(\ref{newtonsinens}), where $n=1$ corresponds to the error function. 
Recall that the Newtonian potential, $\Phi(r)=  -\frac{\mu}{4\pi^2 M^2_p} \frac{f(r)}{r}$. For illustrative purposes, we have taken $M=4\times 10^{-3}$~eV. }
\label{Fig1}
\end{figure}

Note that for an {\it entire function}, we can always treat $f(r)$ as a polynomial function. As a warm-up exercise we note that the sine Fourier transformation for 
\begin{equation}
\tau=-\frac{k^{2n}}{M^{2n}}\,,
\end{equation}
 gives
\begin{equation}
        f(r)=\frac {Mr}{n}\sum_{p=0}^\infty(-1)^p\frac{\Gamma(\frac pn+\frac1{2n})}{(2p+1)!}(Mr)^{2p}\,,
        \label{newtonsinens}
\end{equation}
using the Gamma function $\Gamma(x)\equiv (x-1)!$. The above result is a generalisation of~\cite{Frolov:2015usa}, where the authors have analysed special cases for
$n=1,2,4$. 
From Fig.~[\ref{Fig1}] we see that the Newtonian potential never blows up at $r=0$. 

An important observation here is that by increasing the value of $n$ yields larger modulation for large $r$, giving us a clear deviation from predictions of 
GR at larger distances, and providing us with a glimpse of testing the non-locality scale $M$. 
We can see that by having
higher modes we now switch on a new mechanism that can be falsifiable in
 a near-future experiment.

Tests
of the inverse square-law assume that departure from the Newtonian potential
follows a Yukawa
potential, $V(r)=-V_0 \left[1 + \alpha \exp(-r/\lambda)
\right]$.
In~\cite{Kapner:2006si}, Adelberger et al. found in 2007~\footnote{Although further tests have been carried out, such as in \cite{Tan:2016vwu}, 
none of these give a stronger constraint on $r$ for $\alpha=1$.} that 
this potential was ruled out for $\alpha$=$1$ down to a length scale of $5.6 \times 10^{-5}$m,
 which means that we can now constrain  $M$ for each $\tau(-k^2)$.  

For each specific value of $n$ in Eq.~(\ref{newtonsinens}), we can check for what value of $M$ that our potential would be detectable by~\cite{Kapner:2006si}. 
The experiment ruled out a Yukawa potential $V(r)=V_0/r (1+\exp(-r/5.6 \times 10^{-5})$ down to length scales of $5.6 \times 10^{-5}$m. 
 Since this Yukawa potential is already ruled out, if a particular value of the scale of non-locality $M$
 provides a larger divergence from GR than this potential, then it can also be ruled out, as otherwise it would have been detected by~\cite{Kapner:2006si}. 
 Using Eq.~(\ref{newtonsinens}), this occurs at $M\sim 0.004,~0.02,~0.03,~0.05$~eV
  for $n=1,~2,~4,~8$ so we can set these as our lower bounds on the scale of non-locality.

Clearly this still leaves us with a large hierarchy between $M\geq 0.004$~eV to $M_p\sim 2.4\times 10^{18}$ GeV, which signifies  that indeed very little is known 
about the gravitational interaction.

Now, let us illustrate the most general situation, when $\tau$ is not a monomial, we may represent it as 
\begin{equation}
\tau(-k^2)=- \frac{k^2}{M^2} + \rho(k^2)\,.
\end{equation}
If we expand $e^{\rho(k^2)}= \sum_m \rho_m k^{2m}/M^{2m}$ (clearly $\rho_0=1$), we yield the sine Fourier transformation 
of $e^{\tau(-k^2)}$ 
\begin{equation} \label{eq:fofrintegral}
        f(r) = \sum^\infty_{m=0} \rho_m (-1)^m \frac{\partial^m}{\partial \alpha^m} \int \frac{dk}{k} e^{-\alpha  \frac{k^{2}}{M^{2}}} \sin (k\hspace{0.4mm}r),
\end{equation}
where $\alpha$ is a dummy variable which we later set to 1. We can calculate \eqref{eq:fofrintegral} either explicitly as%
\ba \label{eq:explicitformoffofr}
        f(r) =  \sum^\infty_{m,p=0}\rho_m (-1)^p  \frac{\Gamma(m+p+\frac{1}{2})}{(2p+1)!}  \left(Mr\right)^{2p+1}
,\ea
or using Hermitian polynomials $H_m(x)$ as
\ba \label{newtonsinerhoshermitian}
        f(r)&=& \pi \mathrm{erf} \left(\frac{M r}{2}  \right) \\\nonumber
        &&-2 \sqrt{ \pi}  e^{-\frac{M^2r^2}{4}} \sum^{\infty}_{m=1} \rho_m (-1)^m  \frac{1}{4^m}  H_{2m-1}\left( \frac{M r}{2} \right)\,.       
\ea
Note that Eq.~(\ref{newtonsinerhoshermitian}) converges to a constant if $\rho_m$ decreases at least as fast as $\frac{(-1)^m}{m!}$, i.e. $\rho=-k^2/M^2$. 

In order to satisfy the low energy requirements of the underlying physics,
 we require that the function $e^{\tau(-k^2)}$ falls at least as fast as $e^{-k^2/M^2}$\cite{Biswas:2011ar}. Any $e^{\tau(-k^2)}$ which does this will also fulfil the convergence condition for Eq.~(\ref{newtonsinerhoshermitian}), meaning that any physically realistic $a(\Box)$ will give a Newtonian potential which returns to the GR $1/r$ potential in the IR limit.  

\begin{figure}[!htbp]
                \includegraphics[width=90mm]{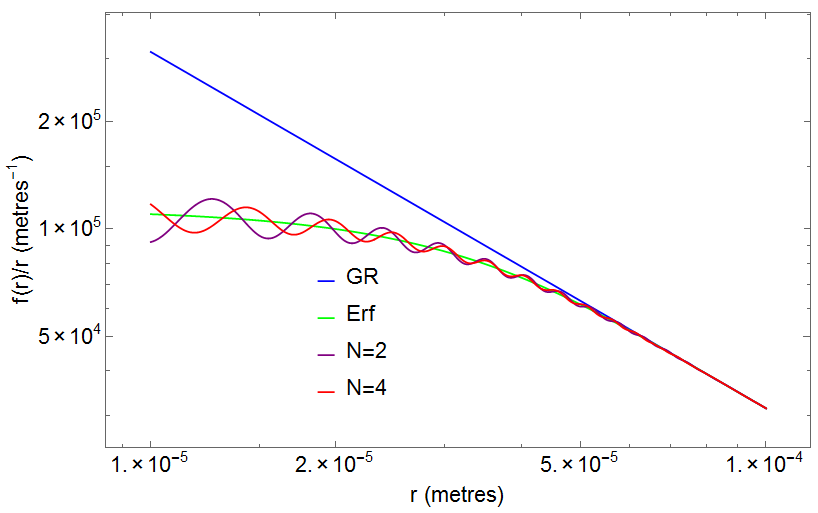}

\caption{We have plotted $f(r)/r$ vs $r$ for Eq.~(\ref{newtonsinerhoshermitian}) and (\ref{eq:nextsimplestcase}), where we have chosen $a_2=4.65\times 10^{-3}$ and $a_4=1.24
\times 10^{-7}$, and for illustrative purposes, we have set $M=4\times 10^{-3}$ eV. }\label{Fig2}
\end{figure}

Next, in order to graphically show the behaviour of Eq.~(\ref{newtonsinerhoshermitian}) in Fig. [\ref{Fig2}], we take the next simplest case, where $\tau$ is the binomial 
\begin{equation}
\tau=-\frac{k^2}{M^2}-a_N \frac{k^{2N}}{M^{2N}}\,,
\end{equation}
and the choice of $a_N$ is motivated by the purpose of illustration of
the oscillations that occur for $r \approx M^{-1}$. In this case, 
\ba \label{eq:nextsimplestcase}
\rho_m=\frac{(-a_N)^{m/N}}{(m/N)!} \text{ for } \frac{m}{N} \in \mathbb{N}
\text{ and zero otherwise. \hspace{1mm}}
\ea\vspace{-2mm}

\section{Conclusion}

Let us conclude by pointing out that {\it infinite derivative, ghost-free} theories of gravity pose a real falsifiable feature compared to GR, which can be tested by measuring the Newtonian potential in near future experiments. We have shown that there exists a universal class of {\it entire} functions for which the theory is {\it ghost-free} as well as 
{\it singularity free} in the UV, while leaving some tantalisingly small effects in the IR, albeit falling as $1/r$-fall of Newtonian potential.
The current experimental limit puts the bound on non-locality to be around
$M\sim 0.004$~eV. Indeed, it is intriguing to reiterate that we know very little about gravity and any modification from Newtonian potential can occur in the gulf of scales spanning some $30$ orders of magnitude, i.e. $0.004~{\rm eV}\leq M \leq 10^{18}$~GeV, but this window also provides an opportunity for testing gravity at short distances.

\section{Acknowledgments}
We would like to thank Tirthabir Biswas, Valeri Frolov,  Saleh Qutub, Terry Tomboulis and Spyridon Talaganis for helpful discussions.
AK is supported by the FCT Portugal fellowship SFRH/BPD/105212/2014 and
in part by RFBR grant 14-01-00707.
 The work of A.M. is supported in part by the Lancaster-Manchester-Sheffield Consortium for Fundamental Physics  
under  STFC  grant  ST/L000520/1.

\end{document}